\documentclass{aa}

\usepackage{graphicx}
\usepackage{txfonts}

\begin{document}
   \title{Spectral and spatial observations of microwave spikes
and zebra structure  in the short radio burst of May 29, 2003}

   \author{G. P. Chernov
          \inst{1,2}
          \and
          R. A. Sych\inst{1,3}
          \and
       N. S. Meshalkina\inst{1,3}
         \and
         Y. Yan\inst{1}
         \and
          C. Tan\inst{1}
          }
   \institute{Key Laboratory of Solar Activity, National Astronomical
Observatories, Chinese Academy of Sciences, 20A, Datun Road,
Beijing 100012, People's Republic of China\\
\email{gchernov@bao.ac.cn}
          \and
             N.V. Pushkov Institute of Terrestrial Magnetism,
             Ionosphere and Radio Wave Propagation
   Troitsk, Moscow region 142190, Russia\\
             \email{gchernov@izmiran.rssi.ru}
             \and
             Institute of Solar-Terrestrial Physics of Siberian Branch
             of Russian Academy of Sciences,
    126a Lermontov Street, Irkutsk, 664033, Russia
             \email{sych@iszf.irk.ru}
                 }

   \date{Received April 06, 2011; accepted May .., 2011}

  \abstract
     {The unusual radio burst of May 29, 2003
connected with the M1.5 flare in AR 10368 has been analyzed. It
was observed by the Solar Broadband Radio Spectrometer
(SBRS/Huairou station, Beijing) in the 5.2--7.6 GHz range. It
proved to be only the third case of a neat zebra structure
appearing among all observations at such high frequencies. Despite
the short duration of the burst (25 s), it provided a wealth of
data for studying the superfine structure with millisecond
resolution (5 ms).} {We localize the site of emission sources in
the flare region, estimate plasma parameters in the generation
sites, and suggest applicable mechanisms for interpretating spikes
and zebra-structure generation.}
 {We analyze of flare area structures and spectral
  parameters of millisecond spikes
and their radio sources. We then interpret the superfine structure
in the framework of known models.}
 {Positions of radio bursts were obtained by
the Siberian Solar Radio Telescope (SSRT) (5.7 GHz) and Nobeyama
radioheliograph (NoRH) (17 GHz). The flare configuration includes
two systems of loops with the common base near the N-spot. The
loop bases coincide with polarized emission sources at 17 GHz. The
sources in intensity gravitated to tops of short loops at 17 GHz,
and to long loops at 5.7 GHz. Short pulses at 17 GHz (with a
temporal resolution of 100 ms) are registered in the R-polarized
source over the N-magnetic polarity (extraordinary mode). The
positions of the subsecond pulse sources at 5.7 GHz change from
pulse to pulse and are level with the tops of some loops over the
magnetic field's neutral line. Dynamic spectra show that all the
emission comprised millisecond pulses (spikes) of 5-10 ms duration
in the instantaneous band of 70 to 100 MHz, forming the superfine
structure of different bursts, essentially in the form of fast or
slow-drift fibers and various zebra-structure stripes. Five scales
of zebra structures have been singled out.} {As the main mechanism
for generating spikes (as the initial emission) we suggest the
coalescence of plasma waves with whistlers in the pulse regime of
interaction between whistlers and ion-sound waves. In this case
one can explain the appearance of fibers and sporadic
zebra-structure stripes exhibiting the frequency splitting.}


\authorrunning{Chernov et al.}
\titlerunning{Zebra structure of May 29, 2003}
\maketitle


\section{Introduction}

Fast solar radio bursts lasting a few milliseconds on the
background of long-term continual radio emission in decimeter and
microwave ranges have already been studied for more than twenty
years (Benz \cite{benz1986}). They are considered to be tied to
the elementary, primary energy release in a flare area (Benz \&
G\"{u}del \cite{benz1987}). The report by Slottje
(\cite{Slottje78}) of fully polarized spikes in a microwave event
was generally considered evidence of species of spikes that are
entirely different from fast bursts at lower frequencies.
Subsequent observations by Stahli and Magun (\cite{Stahli}) have
confirmed the exceptionality of the phenomenon at microwaves. The
typical duration of single spikes are 50 -- 100 ms around 250 MHz,
and it decreases to 10 -- 50 ms at 460 MHz and up to 3 -- 7 ms at
1420 MHz (Benz \cite{benz1986}). Rozhansky et al.
(\cite{Rozhansky}) have gathered all currently available
measurements of spike duration and find a power law with exponent
1.29 for the spectral range 237 - 2695 MHz. This law predicts that
the duration of spikes at f $>$ 4.5 GHz should be less than 2 ms,
which is well below the spectrometer temporal resolution.

Microwave spikes always have a duration $<$ 10 ms and usually
appear in clusters where spikes are randomly distributed in the
frequency-time plane. It is generally agreed that spikes are
non-thermal, coherent emission that is closely connected with the
particle acceleration and energy release in flares, and during
subsequent years interest in researching spikes increased. Benz
(\cite{benz1985}; \cite{benz1986}) described spikes in the
decimeter range at the base of observations with the digital
spectrometer IKARUS (Zurich). In five events, clear time and
frequency profiles were presented; a good correlation was found
with type III bursts and hard X-ray bursts. From the flare
configuration, a typical upper limit is found to the dimension of
spike sources of 200 km. It was proposed that the observed
fragmentation in the radio emission should already occur in the
exciter thanks to the fragmentation of the primary energy release
(Bastian et al. \cite{Bastian98}).

With a diameter of 200 km and for a circular source, the
brightness temperature of spikes is up to $10^{15}$ K. Only a
coherent emission process can reach such intensity. Fleishman and
Melnikov (\cite{fleishman}) propose exhaustive review of all basic
mechanisms of the excitation of spikes and their relevance to the
observed parameters. Because it is based on the observations of
spikes at the harmonic frequencies and the partial registrations
of the radio emission of spikes in the extraordinary mode, they
conclude that the spikes can only be excited by the cyclotron
maser mechanism (the loss-cone instability of the first and second
electromagnetic cyclotron harmonics). The electron cyclotron maser
mechanism is examined in Fleishman and Melnikov (\cite{fleishman})
in detail. The basic objection against this mechanism comes from
many authors` estimations of the absorption of radio emission at
the third gyro-resonance level in the corona. Furthermore, this
mechanism is only effective in those sources where the ratio of
plasma and cyclotron frequencies is less than unity ($\omega_{\rm
pe}/{\omega_{\rm Be}} < 1$), while practically all observational
data provide evidence of the inverse relationship ($\omega_{\rm
pe}/{\omega_{\rm Be}} \gg 1$), not only in the meter but even in
the microwave range.

Fleishman et al. (\cite{fleishman2003}) only propose indirect
evidence of the ratio $\omega_{\rm pe}/{\omega_{\rm Be}} \leq 1$,
related to the flat spectrum of gyrosynchrotron emission. This
means that no indication of the Razin effect in the
spike-producing bursts is present because the Razin effect begin
to suppress the low-frequency part of the spectrum for the ratio
$\omega_{\rm pe}/{\omega_{\rm Be}} > 1$. However a contribution
into low-frequency part can give another radiation mechanism (for
example, plasma emission). Besides this, some important properties
of spikes have not been taken into account by Fleishman \&
Melnikov (\cite{fleishman}), and Chernov et al.
(\cite{chernov2001}) considered an alternative generation
mechanism based on the interaction of Langmuir waves with
ion-sound waves.

Analyzing the burst radio emission fine structure enables us to
identify acceleration mechanisms of fast particles and their
propagation in a flare area. Such investigations become effective
if the positions and sizes of radio sources and radio emission
polarization are measured concurrently with observations of the
dynamic spectrum. Then we can estimate the radiating mode and
generation mechanisms (Meshalkina et al. \cite{meshalkina}). The
sources of subsecond pulses can be localized, if they occur in the
operating frequency bands of the Siberian Solar Radio Telescope
(SSRT) at about 5.7~GHz. Such a set of observations was carried
out on May 29, 2003 for a small microwave burst of only $\sim$ 20
s duration, but enjoying more variety in the fine structures. The
burst was remarkable for all its parameters: all emission
comprised millisecond spikes, for ten seconds the spikes grouped
in the form of fast-drift bursts, slow-drift fibers, and
zebra-structure (ZS) stripes on five scales according to the
frequency separation and stripe drift. The radio emission
polarization was weak, but had different signs in different
structures and changed with time. Such a combination of fine and
superfine structures is of great interest to check known radio
emission generation mechanisms.

   \begin{figure*}
   \centering
   \includegraphics[width=13cm]{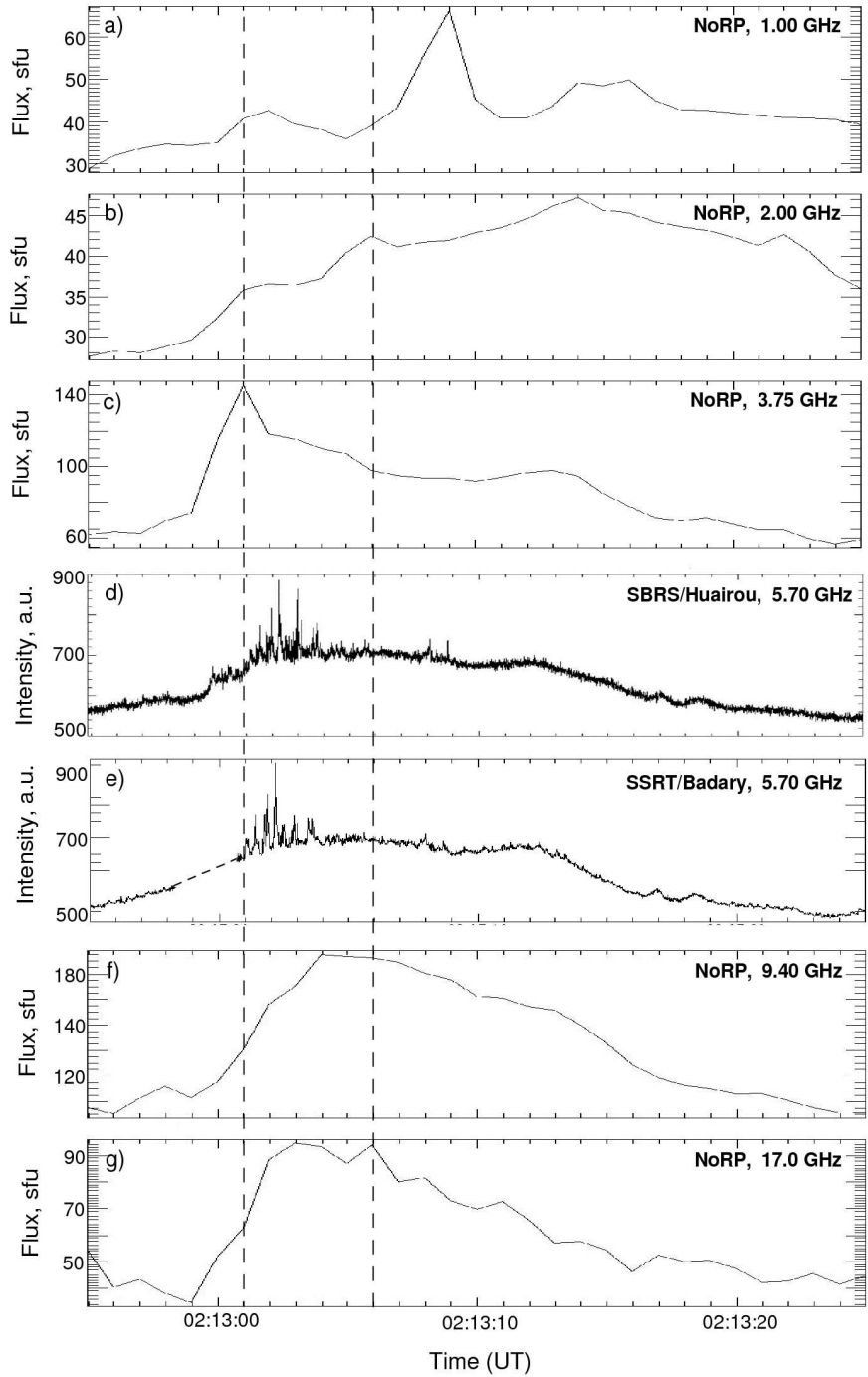}
      \caption{The intensity profiles recorded by Nobeyama spectropolarimeters (a--c, f , g),
       SBRS (d), SSRT (e)
      (including the quiet Sun level) in the interval 02:12:55--02:13:25 UT.
      The vertical dashed lines mark the interval with the fine structure at 5.7
GHz.).
              }
         \label{lfig1}
   \end{figure*}

We have previously described several phenomena in which
millisecond spikes made up the superfine structure for other
elements in the fine structure (e.g. ZS-stripes) (Chernov et al.
\cite{chernov2003}, \cite{chernov2006b}). But at that time, these
were long-term events in which the conditions for forming such a
superfine structure were being evolved over several hours, whereas
this short-term phenomenon exhibited all the variety for 5~s, when
it was immediately obvious that all the emission consisted of
millisecond spikes. And, as opposed to many other events, here we
can register the position of radio sources. Thus, in deciding on a
model, it is necessary to consider this main circumstance. We
should based it on the condition that the emission of millisecond
spikes is primary, and in one source all the variety of fine
structures in the form of fibers and zebra stripes develop over
several seconds.

Spectral data and time intensity profiles from different
observatories are discussed in Section 2. SSRT data on the
positions of radio sources at 5.7 GHz are compared with the
Nobeyama radioheliograph data at 17 GHz. Section 3 discusses
observations and possible radio emission mechanisms.

\section{Observations}
\subsection{Instruments}

The extraordinary phenomenon of May 29, 2003 was discovered in the
dynamic spectra of the Solar Broadband Radio Spectrometer (station
Huairou) in the frequency band of 5.2 to 7.6 GHz. 120 frequency
channels in this band (every 20~MHz) provide the 5 ms temporal
resolution in the right (R) and left (L) circular polarization (Fu
et al. \cite{fu}).

We used microwave total flux records from Nobeyama Radio
Polarimeters NoRP, (Torii et al. \cite{Torii}; Shibasaki et al.
\cite{Shibasaki}; Nakajima et al. \cite{Nakajima1985}) at 1, 2,
3.75, 9.4, and 17~GHz. The time resolution of routine NoRP data
available at the NoRP Web site is 1 s, and for flares it is 0.1 s.
We used the images obtained by Nobeyama Radioheliograph (Nakajima
et al. \cite{Nakajima1994}) at 17 (I, V) (temporal resolution is
0.1 s). The spatial resolution is 10 $\arcsec$.

Positions and sizes of the radio sources were determined from
observed data obtained at the SSRT, cross-shaped
radiointerferometer operating in the 5.67--5.79 GHz frequency
range (Smolkov et al. \cite{Smolkov}; Grechnev et al.
\cite{grechnev}). Two-dimensional solar disk maps are acquired
every two to three minutes depending on the solar declination and
the hour angle of observations. The full temporal resolution of
one-dimensional scans is $\sim$ 14 ms at the spatial resolution of
$\sim$ 15$\arcsec$. Time profiles of intensity (I = R + L) and
polarization (V = R - L) of fast bursts are regularly put into the
SSRT database on the website
http://badary.iszf.irk.ru/Ftevents.php. Widths of one-dimensional
SSRT diagrams, when the radio burst was recorded, were 16$\arcsec$
for the north--south antenna beam and 21$\arcsec$ for the
east--west beam. Inclination angles to the central meridian of the
Sun were -36$^{\circ}$ and 39$^{\circ}$, respectively.

\subsection{Observed data}

\begin{figure*}
        \centering
        \parbox{0.9\hsize}{
    \resizebox{
 \hsize}{!}{\includegraphics[width=10cm,clip=,
                 bb=1 1 550 651]{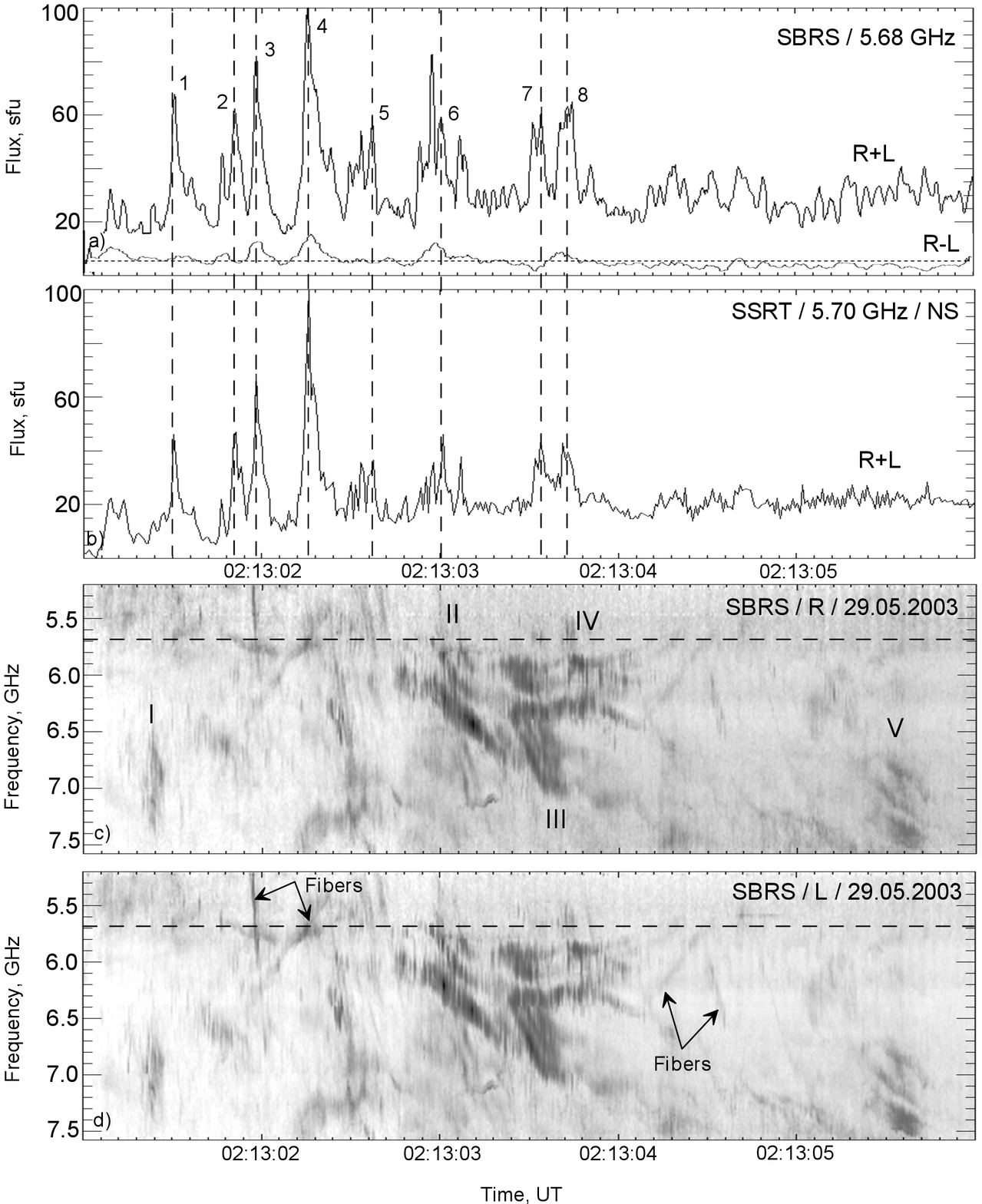} }}
      \caption{Comparison of the time profiles recorded at SBRS
      (a) and SSRT (b)
       and the dynamic spectrum (c, d) registered by the SBRS spectropolarimeter.
       The dark details
correspond to the increased emission. The horizontal dashed line
(c) denotes the SSRT receiving frequency at 5.7 GHz.
      The calibration was performed using the SSRT-registered background burst.
      Arabic
numerals (a, b, dashed vertical lines) mark the maximum moments of
subsecond pulses recorded by the SSRT.   }
         \label{lfig2}
   \end{figure*}

The May 29, 2003 short radio burst of $\sim$ 20 s duration in the
main phase (02:13:00--02:13:20 UT) was related to the flare in the
southen active region 10368 (S37E03), M1.5/1F in the GOES
classification (02:09--02:24 UT). It occurred during the decay
phase of a larger flare (X1.2) with a maximum at 01:05 UT, tied to
another region 10365 (S06W37).

Figure \ref{lfig1} gives a general picture of this phenomenon at
five frequencies of the Nobeyama spectropolarimeters and
simultaneous observations at the SSRT and the SBRS
spectropolarimeter at 5.7 GHz. The total duration of the radio
burst was about 1.5 min. The maximum flux ($\sim$ 180 sfu) was
registered at frequencies around 9.4 GHz  (1 sfu =$ 10^{-22}
$Wm$^{-2}$Hz$^{-1}$). It shows the interval with the fine
structure at 5.7 GHz, located just between 3 and 9 GHz maxima
(including all ZP stripes). It is seen that the maximum of
background burst spectrum is located near 9 GHz, and it is late
for three seconds relative to maximum on 3.75 GHz (the intensity
scale for NoRP is in solar flux units (sfu).

Figure~\ref{lfig2} presents the intensity (Stokes parameter I) and
polarization profiles (Stokes parameter V, lower curve) of the
SBRS spectropolarimeter at 5.68 GHz (a) and the SSRT (Stokes
parameter I, the NS line)(b) and the dynamic spectra in the left
and right polarization of the SBRS spectrometer in the 5.2--7.6
GHz frequency band (c, d) in the interval marked off in
Fig.~\ref{lfig1}. To remove the hardware interference we prepared
a wavelet cleaning of the dynamic radio spectrum (Sych \& Yan
\cite{Sych}) with deleted high-frequency noise around 13 Hz along
the frequency channels. This technique allows the original solar
signals from noise component to be separated.

The continual emission at 5.7 GHz was weakly polarized (the left
sign within the limits of errors made up 5\%). Nearly all the fast
bursts were right hand circularly polarized ($\sim$ 20\%). The
amplitude of RCP and LCP at the dynamic spectra in
Figs.~\ref{lfig2}c, d are identical, which points to a weak
polarization degree, which is clearly visible on the ratio between
polarization and intensity curves (Fig.~\ref{lfig2}a). This
behavior of polarization is characteristic of the entire range of
5.2 to 7.6 GHz.

The emphasis will be on analyzing the fine structure in the
interval 02:13:00--02:13:06 UT. The similarity between amplitudes
and shapes of the subsecond pulses measured independently by two
spaced instruments indicates that the observed fine time structure
has a solar origin. The amplitude of the subsecond pulses is
relatively high and reaches up to 60 sfu.

\begin{figure*}
\centering
\parbox{0.9\hsize}{
\resizebox{
 \hsize}{!}{\includegraphics[width=13cm,
                 bb=1 1 548 549]{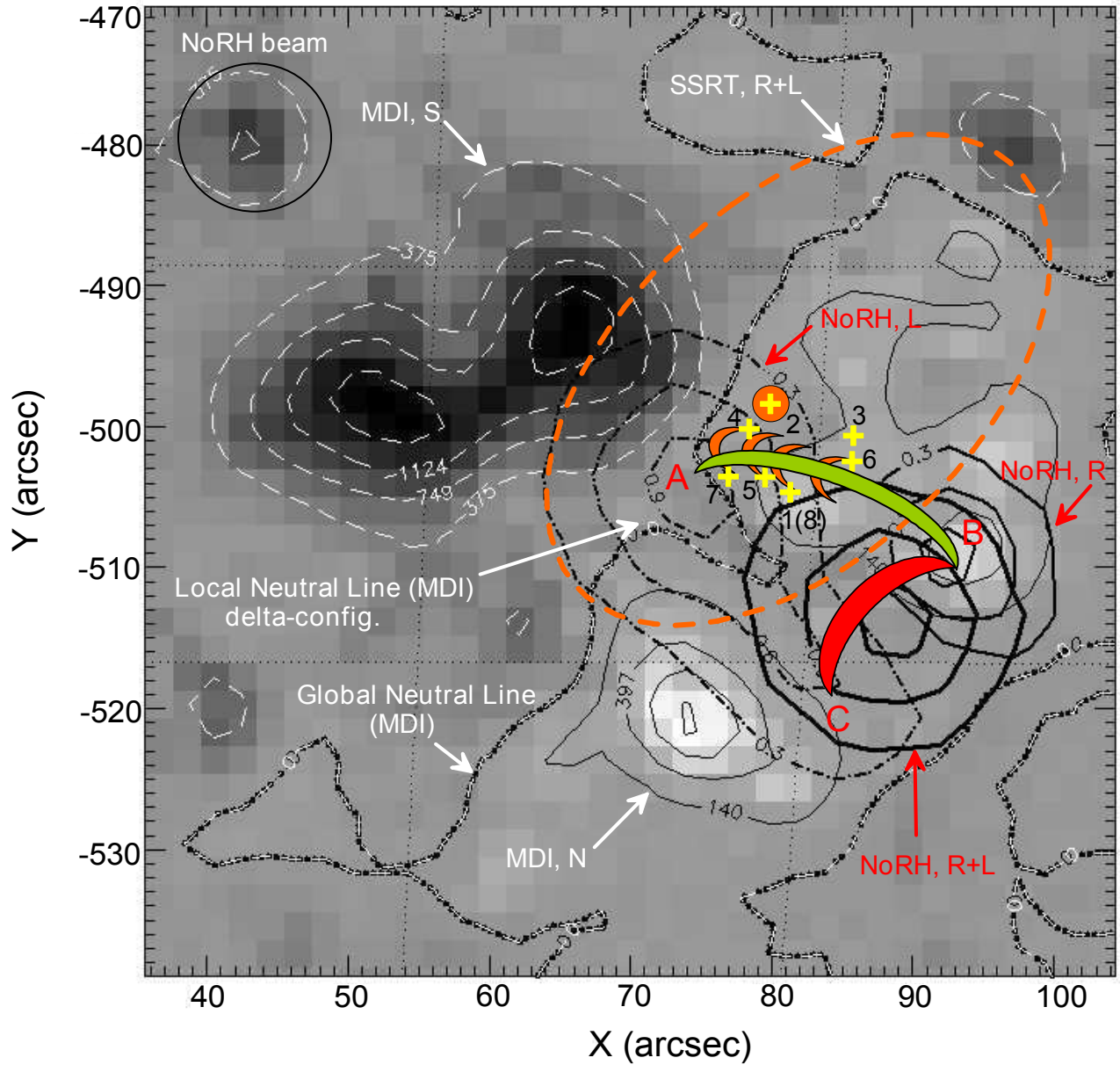}} }
      \caption{The positions of radio sources at 5.7 GHz (SSRT)
       and 17 GHz (NoRH) superimposed on the MDI magnetogram.
       The orange dashed line indicates the position of the left
       polarized local source at 5.7 GHz at 02:17:54.8 UT.
       The dash-and-dot line shows the position of the
        left polarized background source of the burst at 17 GHz
        (extended southwestward) at 02:13:00 UT. The straight
        yellow crosses are the positions of spike sources 1--8
        corresponding ones in
        Fig.~\ref{lfig2}.
              }
         \label{lfig3}
   \end{figure*}

Various structures with relatively narrow instantaneous spectrum
bands were observed in the background of the continuum emission
(Figs.~\ref{lfig2}c, d). Among them are some fibers whose emission
frequency varied with time. The spectrum also has several zebra
structures separated into intervals.

It is worth noting that pulses 1 and 4 correspond to the
intersection of the SSRT receiving band by bursts that drift to
low frequencies in the dynamic spectra (and pulse 2 - to high
frequencies); pulses 3 and 5 -  to fibers, which drift in the
opposite direction and look like fiber bursts. The sixth pulse is
the superfine structure of the zebra-stripe, low-frequency edge.
Pulses 7 and 8 are close to the horizontal zebra-structure stripe
in episode IV; here, pulse 7 has inverse polarization.

Spectra in Figs.~\ref{lfig2}c, d demonstrate that all the emission
comprises fine-structure details in the form of pulses (spikes).
Their duration is limited to one pixel; i.e., it is at the limit
of the 5 ms temporal resolution (rarely to two pixels, $\sim$ 10
ms). They occupy the frequency band in the spectrum of $\sim$ 70
to 100~MHz. Their band and duration might be below the resolution
limit. The continual background burst also exhibits the spike
structure. Groups of spikes create various fine-structure forms:
fast and slow-drift fibers. The brightest (most intense) ones form
different zebra-structure stripes.

At 17 GHz, all subsecond pulses were only registered in the
R-channel when the L-polarity emission background increased. The
correlation curve of the burst in the polarization channel
displays the origin of the fine temporal structures (pulses) with
300 ms periodicity. We can consider taking sensitivity curves into
account for the appearance of fine spatial details about small
angular size of pulse sources ($< 20 \arcsec$). To verify the
parameters of observed R-polarity pulses during the maximum
activity (02:13:00-02:13:06 UT), we synthesized the radio images
at 17 GHz with 100 ms cadence, prepared the wavelet spectrum of
amplitude profile for the new R-source, and constructed global
wavelet spectrum (GWS-spectrum). This spectrum is shown for the
SSP-related periods $\approx$ 300-400 ms. Comparing those values
with the obtained periods of intensity oscillations in the SBRS
dynamic spectrum (5.2--7.6 GHz) shows the presence of maxima
oscillation in this period range. We can suppose that these
periods can be considered as chains of the spikes in SBRS spectra.

\subsection{Zebra structure}

The zebra-structure fragments are shown in the lower panels (c, d)
of Fig.\ref{lfig2}. Drift ZS stripes of different frequency scales
were observed for four seconds in a wide frequency range (5.8--7.2
GHz). All the stripes were made up of isolated pulses. The
shortest ZS episode I (around 7 GHz) did not contain clear
stripes. Such ZS behavior is known in the literature as "braided"
ZS (Slottje \cite{Slottje81}). In episode II, the frequency drift
of stripes $df/dt \approx$ 1.3--1.5 GHz s$^{-1}$ with a frequency
separation between the stripes $\Delta{f} \approx$ 300--350 MHz.
In episode III, $df/dt \approx$ 3.2-3.6 GHz s$^{-1}$ with
$\Delta{f} \approx$ 170--200 MHz. From approximately 02:13:03.3
UT, two neat almost stationary stripes (IV) with a frequency
separation of 450 MHz appeared between 5.8 and 6.4 GHz. Fragments
of slow-drift stripes are visible between them. After 02:13:03.8
UT they might have been taken for the frequency splitting of two
main stripes. Nearly 1.5 s after this fragment, in the interval of
02:13:05.4 to 02:13:05.8 UT at 6.6--7.6 GHz, there were three more
ZS stripes (V) with $df/dt \approx$ 1500 MHz/s and $\Delta f
\approx$ 250 MHz. They were weak and diffuse, but had a spike
structure and occurred with additional fragments between stripes
(frequency splitting). There was also a very short fragment of
small-scale ZS at 02:13:03.3 UT at $\sim$ 7.150 GHz: the frequency
separation was $\sim$ 40 MHz; the width of these two stripes in
the emission at the limit of the instrument resolution $\sim$ 20
MHz.

At lower frequencies on SBRS spectra, 2.6--3.8 GHz, there was no
ZS at that time (02:13:00--02:13:05 UT). Notable there are only
several groups of drift spikes of $\sim$ 20-30 ms duration each.
The pattern of activity differed markedly from what is described
in the range of 5.2 to 7.6 GHz. The main difference is associated
with the lack of many pulses. Though the majority of bursts of
200--300 ms duration contain some inhomogeneity, they do not
exhibit the superfine spike structure.

In the decimeter range (1--2 GHz), there were no spikes. The patch
bursts were even more prolonged ($>$ 0.5 s) and, drifting
gradually to high frequencies, formed a structure resembling a
type II burst in its initial frequencies.

\subsection{The flare area structure}

The configuration of flare structures in the active region has
been investigated using the time series of radio maps in intensity
and polarization obtained at NoRH at 17 GHz. During the flare two
sources of polarization were revealed: a compact source with right
hand polarization in the west (R)(marked as B in Fig. \ref{lfig3})
near the N-spot and a region in the east, extended from north to
south with oppositely directed circular polarization (L). This
region exhibited two centers of brightness, on the $L_N$ and $L_S$
edges (marked as A and C in Fig. \ref{lfig3}). The $L_S$ source
position corresponded to a small isolated N-polarity region; the
north $L_N$ source appeared near the S-polarity spot. The
combination of all the data on the positions of radio sources is
presented in Fig. \ref{lfig3}.

The flare configuration may be expected to be based on two loop
systems $R-L_S$ (long) and $R-L_N$ (short). The presence of a
shorter loop with bases in $R$ and $L_S$ polarized emission
centers is confirmed by the location of emission intensity center
at 17 GHz between them. Between the bases of the second, longer,
and higher loop system, there is a flare emission region at 5.7
GHz, extended along the photospheric field neutral line.

\begin{figure}
   \centering
   \includegraphics[width=8cm]{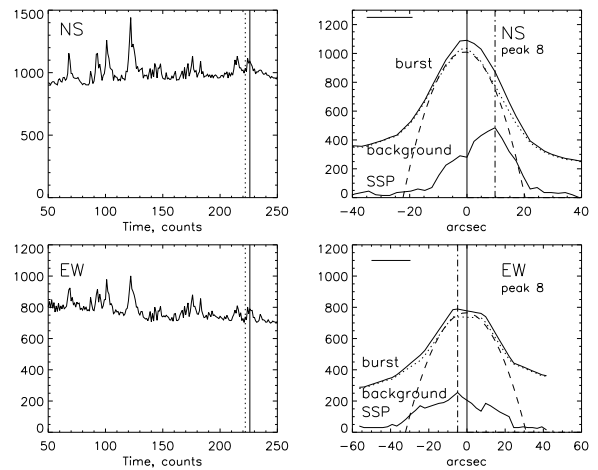}
      \caption{Left: The time intensity profiles obtained from one-dimensional scans.
Right: The one-dimensional NS and EW scans during the pulse 8:
positions of the spike sources in intensity distributions relative
to the background burst source (dotted line). The background burst
scan was approximated by the Gaussian distribution (dashed line)
to determine accurately the position of the burst maximum. The
horizontal bar at the top shows the size of the point source at
the spatial half-width. Profile magnitudes are in arbitrary units.
              }
         \label{lfig4}
   \end{figure}

The background in Fig.\ref{lfig3} is the MDI magnetogram taken for
the later period in order to show the evolution of the field
``delta"-configuration (Kunzel et al. \cite{Kunzel}; Bray and
Loughhead \cite{bray}; Severny \cite{Severny}). The thin black
lines (dashed lines are the S-polarity, solid lines the
N-polarity) indicate magnetic structures; the wide stripe is the
neutral line. The coordinates of spike source centers were
obtained using the projection method by comparing one-dimensional
NS and EW scans (Fig.\ref{lfig4}) with the maximum accuracy of 5
to 10$\arcsec$.

The spikes were weakly polarized for the most part, and so there
is no point in determining the wave type. Even for spikes 3 and 4
with moderate polarization (20--30\%), the wave type determination
encounters difficulties in accurately assessing the magnetic field
polarity.

Analysis has revealed that there were three emerging magnetic
structures during the field ``delta"-configuration evolution. They
are lettered A, B, and C. The positions of SSRT spikes group
together just in the place where structure A emerged. All fast
bursts at 17 GHz are located at structure B. The large stripes
(green and red) denote possible flare loops. In their common base,
the highest magnetic field intensity near the photosphere was
$\sim$ 800~G. The main process associated with the occurrence of
R-polarized pulses at 17 GHz is the appearance of a new R source.

The positions of the spike source centers at the SSRT receiving
frequency were estimated from the projection profiles (scans) of
sources shown in Fig. \ref{lfig4}. The SSRT diagram did not
exhibit broadening of the spike sources, therefore their sizes
might be assessed as $<$15$\arcsec$.

\begin{figure}
   \centering
   \resizebox{
 \hsize}{!}{\includegraphics[width=8cm,clip=,
                 bb=1 1 545 575]{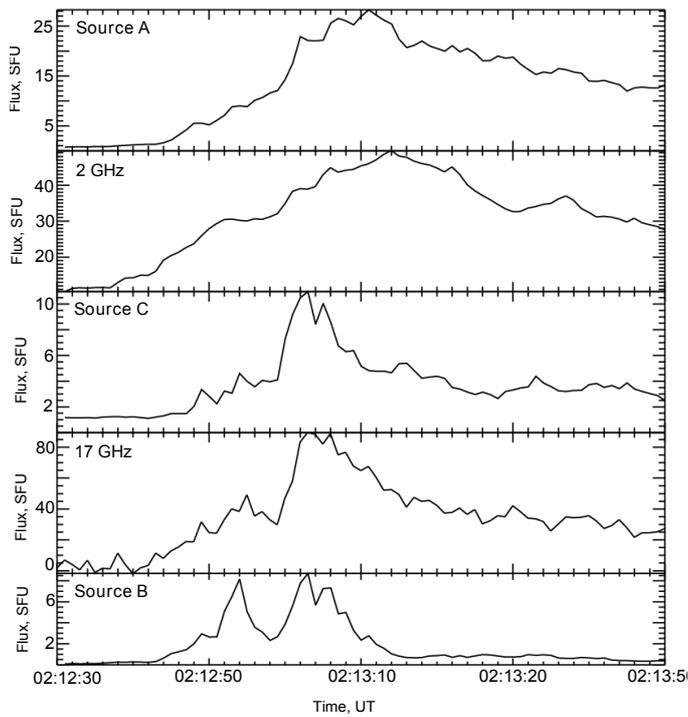} }
      \caption{The comparison of the time profiles of the individual
      sources A, B, C
      at 17 GHz with integral spectra.
It is obvious that the source A emission profile is close to the
integral emission profile at 2 GHz; source C, to that at 17 GHz.
The intensity scale is in solar flux units (SFU).
              }
         \label{lfig5}
   \end{figure}

The assumptions about the magnetic loop structure are confirmed by
comparison of the time variation of emission fluxes (Fig.
\ref{lfig5}) from the sources marked as A, B, and C in Fig.
\ref{lfig3}. It is worth noting that the emission from the short
loop exhibits a higher variability on time scales to 0.1 s - the
temporal resolution of observations at 17 GHz (positions of short
pulse sources). In integral spectra at low frequencies, the main
contribution is made by long loops; at high frequencies, by short
loops. Comparison of polarization signs with directions of the
magnetic field in the sources points to the emission of an
extraordinary wave in the background flare burst that conforms to
the gyrosynchrotron emission mechanism (Fig.~\ref{lfig3}).

\subsection{Brief conclusions from the observations}
 \begin{enumerate}
      \item The flare consists of two loop systems with close N-bases
       and spaced S-bases (flare of type Hanaoka (\cite{hanaoka})).
       The flare initiation is associated with the dynamics of
       the isolated S-region, the formation of which led to
       the ``delta"-configuration of the flare area.
      \item The microwave emission spectra in both loops
      suggest the gyrosynchrotron nature of the emission
      with frequencies of the spectra maxima as evidence
      that in short loops there is a larger magnetic field
      and a greater number of accelerated electrons.
    \item The sources of R-polarized pulses at 17 GHz were
    located in the N-base of the long flare loop, and
    narrow-band structures at 5--7 GHz at the tops of
    the long loops over the neutral line.
    \item Fast pulses (spikes) of 5--10 ms in the
    instantaneous frequency band 70--100 MHz are superfine
    structures of various drift bursts, fibers, and
    zebra-structure stripes.
    \item Most pulses (in the spectrum 5.2--7.6 GHz) are
    moderately R-polarized.
    \item The zebra structure appeared about 3 s after
    the first spikes in the form of five patterns of
    stripes on different scales (in the frequency drift,
    stripe separation). Some stripes bifurcate
    (exhibited the frequency splitting).
    \item The significant difference between parameters
    of each zebra-structure pattern suggests that
    their sources resided in different magnetic loops
    (different footpoints or tops of the loops).
   \end{enumerate}

\section{Discussion of observations and possible mechanisms}

On May 29, 2003, about six hours before the flare under study in
AR 10368, the process  began of long-term magnetic field
reconfiguration and development of a potentially hazardous (in
terms of flare initiation) ``delta"-configuration of the magnetic
field. New magnetic fields emerged from under the photosphere, and
local neutral lines appeared.

The onset of the background burst must have been tied to the onset
of new magnetic flux emergence. Two new flare loop complexes were
formed from new emerging magnetic structures related with R and L
radio sources at 17 GHz. The position of the centers of the
sources of spikes 1--8 indicates the electron fluxes in long
loops.

The dynamic spectrum in the 5.2--7.5 GHz frequency band attests to
the presence of several sources with similar physical conditions,
and the positions of their sources allow us to assert that they
are located near the tops of long magnetic loops. The frequency
drift is directed predominantly to high frequencies that might be
connected with the gradual density increase in density of flare
loops during the flare. The density increases at different rates
in different loops. The emission of narrow structures is formed by
pulses with a frequency band of about 100 MHz and a duration of
5--10 ms.

The frequency drift of burst maximum to the high frequencies
(Fig.~\ref{lfig1}) is possibly related to the motion of the
accelerated particles downward. The background continuum burst is
similar to type III burst with the reverse drift, and we analyze
fast millisecond fine structures against its background.

\subsection{Generation mechanisms}

It is natural to attribute the weak emission of the entire
background burst of about 25 s to the gyrosynchrotron emission of
nonthermal electron fluxes with a wide energy spectrum (100--200
keV) flowing from a reconnection region down toward a transition
layer. The R-polarization subsecond pulses at 17 GHz may be
induced by more energetic quasi-periodic particle beams (Altyntsev
et al. \cite{altyntsev2000}). The emission of fine structures in
dynamic spectra is associated with coherent mechanisms.

The mazer mechanism (ECME, Fleishman \& Melnikov,
\cite{fleishman}) drops out because of the low-intensity magnetic
field under the region of the sources ($<$ 200~G). In this case,
the cyclotron harmonics above the tenth one become ineffective.

The plasma mechanism (beam instability) is quite possible (Zaitsev
et al. \cite{zaitsev}). It provides the millisecond duration for
radio brightness pulses in a microwave range. This mechanism may
first be associated with the reverse drift bursts (at 02:13:02
UT).

The interaction between plasma waves and whistlers in a pulse
regime and the interaction between whistlers and ion-sonic waves
are also distinct possibilities (Chernov et al.
\cite{chernov2003}; Chernov et al. \cite{chernov2001}). The
presence of whistlers in flare loops is confirmed by the
appearance of fibers and zebra-structure stripes. This mechanism
may even be preferable because it can explain the observable
chaotic distribution of structures in the wide frequency range
(all the emission consisted of spikes).

In this case, one can naturally explain the appearance of fibers
and zebra structure by the same superfine structure in the form of
millisecond pulses (Chernov \cite{chernov2006}). The whistlers
were gradually captured into the flare loops, and different loops
gave stripes on different scales. The instantaneous emission band
of an isolated pulse is determined from linear sizes of particle
beams with small velocity dispersion (beam instability) or from
sizes of whistler wave packets (in a model with whistlers).

\subsection{ZS model at DPR}

The widespread mechanism of zebra structure under double plasma
resonance (DPR) conditions (Zheleznyakov \& Zlotnik
\cite{Zheleznykova}, \cite{Zheleznykovb}; Winglee \& Dulk,
\cite{winglee}) is engaged when, at discrete levels, in the
magnetic trap the upper hybrid frequency equals to an integer of
electron cyclotron harmonics. This phenomenon may be used as an
example to demonstrate the possible difficulties that the
mechanism at DPR faces for explaining the zebra structure with
sporadic stripes. Originally, it was developed on the condition
that there are always DPR levels in the source, and only the
presence of fast particles with loss-cone or ring (DGH) velocity
distribution provides the excitation of the regular zebra
structure of different modulation depths (depending on the
particle energy spectrum) (Kuznetsov \& Tsap
\cite{kuznetsovtsap}). The presence of such particles may be
considered obvious, so that the absence of zebra-structure at the
beginning of the burst suggests there are no DPR levels.

DPR levels cannot quickly appear and disappear in the corona.
Processes of emergence of new magnetic fluxes and formation of new
loops are very slow. Changes in height scales of density and
magnetic fields may be associated only with flare ejections and
shocks. With the ejection velocity taken as Alfv\'{e}n velocity
($\approx$ 1000 km s$^{-1}$), the DPR levels (in view of the
magnetic field line freezing-in) may shift the flare loop by half
(green in Fig. \ref{lfig3}) with sizes of $\sim 10 \arcsec$ for
approximately 7 s. This time exceeds the lifetime of
zebra-structure stripes by an order of magnitude. The sporadic
nature of ZS stripes, the wave-like drift of two stripes, their
splitting, and the presence of isolated fibers with forward and
reverse drifts rule out the existence of any regular DPR levels.
Moreover, the validity of the DPR conditions in fine flare loops
along which density and magnetic field strength vary only slightly
is in doubt (Aschwanden \cite{aschwanden}). At the same time, all
the effects can be naturally explained in the joint model of zebra
structure and fibers when plasma waves interact with whistlers
(Chernov \cite{chernov2006}).

The unique data on the position of radio sources of certain ZS
stripes in the dm-range was obtained the first time in the recent
paper by Chen et al. (\cite{chen}). The comparative analysis of
DPR and whistler models made by the authors should be verified by
new, more accurate estimations based on real plasma parameters.

\subsection{The model with whistlers}

All papers discussing the DPR mechanism (and its upgrading) omit a
very important feature of the fast particle distribution function
with loss cone: the particles excite whistlers simultaneously, and
all the dynamics of the distribution function and emissions depend
on the interaction of fast particles with whistlers (Chernov
\cite{chernov1996}). Let us note first of all that whistler
excitation growth rates in the microwave range exceed their value
in the meter range (Yasnov et al. \cite{yasnov}). According to the
model described in Chernov (\cite{chernov2006}) (Sect. 3.6),
zebra-structure stripes are associated with the whistler
excitation by the anomalous Doppler effect at the top of the
magnetic particle trap. Whistler wave packets fill the magnetic
trap and propagate downward (at a significant angle to the
magnetic field) at a group velocity of $\sim 10^9 $cm s$^{-1}$.
The wave packet size in the corona is determined by the linear
size of relaxation of the whistler-exciting fast particle beam
(Formula (13.4) in Breizman (\cite{breizman})), see also the
discussion of Formula (29) in Chernov (\cite{chernov2006},
\cite{chernov2010}).

In the recent short critical review by Zlotnik (\cite{Zlotnik}),
the advantages of the DPR model and the main failures of the model
with whistlers are refined. The author asserts that the theory
based on the DPR effect is the best-developed theory for a zebra
pattern origin at meter-decimeter wavelengths at the present time.
It explains the fundamental ZP feature in a natural way, namely,
the harmonic structure (frequency spacing, numerous stripes,
frequency drift, etc.) and gives a good fit for the observed radio
spectrum peculiarities with quite reasonable parameters of the
radiating electrons and coronal plasma.  The statement that the
theory based on whistlers is only able to explain a single stripe
(e.g., a fiber burst) was made in Zlotnik (\cite{Zlotnik}) without
the correct ideas of whistler excitation and propagation in the
solar corona.

Zlotnik uses the term "oscillation period" of whistlers connected
with a bounce motion of fast particles in the magnetic trap.
Actually, the loss-cone particle distribution is formed as a
result of several passages of the particles in the magnetic trap.
Kuijpers (\cite{Kuijpers}) explained the periodicity of fiber
burst using this bounce period ($\sim 1$ s). And if we have one
fast injection of fast particles, whistlers (excited at normal
cyclotron resonance) are propagated towards the particles, they
disperse in space. Quasilinear effects therefore do not operate in
normal resonance. ZP is connected instead with whislers excited at
anomalous resonance during long-lasting injection. In such a case,
waves and particles propagate in one direction, quasilinear
effects begin operating, and their role increases with an
increasing duration for injections. ZP is excited because the
magnetic trap should be divided into zones of maximum
amplification of whislers, separated by intervals of whistler
absorption (see more details Chernov \cite{chernov1990}). The
bounce period does not interfere with this process, but it can be
superimposed on ZP.

However, the whistler amplification length is always small (on the
order of $\leq 10^8$ cm in comparison with the length of the
magnetic trap being $> 10^9$ cm) for any energy of fast particles
(Breizman \cite{breizman}; Stepanov and Tsap \cite{Stepanov}).
According to Gladd (\cite{Gladd}), the growth rate of whistlers
for relativistic energies of fast particles decreases slightly if
the full relativistic dispersion is used. In this case, the
whistlers are excited by anisotropic electron distributions due to
anomalous Doppler cyclotron resonance.

Later, Tsang (\cite{Tsang}) specified calculations of relativistic
growth rates of whistlers with the loss-cone distribution
function. It was shown that relativistic effects reduce growth
rates slightly. According to Fig. 8 in Tsang (\cite{Tsang}), the
relativistic growth rate is roughly five times less than the
nonrelativistic growth rate. However, the relativistic growth
rates increase with the perpendicular temperature of hot
electrons. According to Fig. 5 in Tsang (\cite{Tsang}), the growth
rate increases about two times when the electron energy is
increased from 100 to 350 keV, if only to keep fixed other
parameters of hot electrons: loss-cone angle, ratio of
gyrofrequency to plasma frequency, temperature anisotropy
($T_\perp/T_\parallel = 3$).

Thus, it was already known long ago that the whistlers can be
excited by relativistic beam with loss-cone anisotropy. Formula
13.4 in Breizman (\cite{breizman}), used as formula (29) in
Chernov (\cite{chernov2006}) for evaluating the smallest possible
relaxation length of beam, has no limitations in the value of
energy for fast particles. Critical comparison of models has been
repeated in Zlotnik (\cite{Zlotnik}), only with a new remark
concerning the Manley-Rowe relation for the brightness temperature
of electromagnetic radiation in result of coupling of Langmuir and
whistler waves:
$T_b=\frac{{\omega}T_l{T_w}}{{\omega}_l{T_w}+{\omega}_w{T_l}}$.
She states that, since ${\omega}_w \ll {\omega}_l$, in the
denominator, only the first term remains and $T_b$ depends only on
$T_l$, and $T_b \sim T_l$; i.e., the process does not depend on
the level of whistler energy\footnote{This approval was also used
by Altyntsev et al. (\cite{altyntsev2011}), to reject the whistler
model for ZS in this event}.

However, Kuijpers (\cite{Kuijpers}) (formula (32) in page 66)
shows that the second term ${\omega}_w{T_l}$ should be $\gg
{\omega}_l{T_w}$ because $T_l \gg T_w$. An analogous conclusion
was made by Fomichev and Fainshtein (\cite{fomichev}) with a more
exact relation to three wave intensities (used by Chernov and
Fomichev \cite{chernovfom}, see also formula (11) in Chernov
\cite{chernov2006}). Therefore $T_b$ in the process $l + {\omega}
\rightarrow t$ depends mainly on $T_w$. Our conclusion is that the
entire magnetic trap can be divided into intermittent layers of
whistler amplification and absorption remains valid for a broad
energy range of fast particles. In Zlotnik (\cite{Zlotnik}) the
main matter that is ignored is that the model involves quasilinear
interactions of whistlers with fast particles, allowing one to
explain all the fine effects of the ZP dynamics, mainly the
superfine structure of ZP stripes and the oscillating frequency
drift of the stripes, which occurs synchronously with the spatial
drift of radio sources.

For the central ZS frequency of 6.4 GHz, the ratio of the
gyrofrequency to the plasma frequency
$\frac{{\omega}_B}{{\omega}_p} \sim$ 0.1 and the ratio of the cold
plasma concentration to the hot particle one $\frac{n^c}{n^h} \sim
10^6$, we obtain the linear size of relaxation of $\sim
2.8\times10^6$ cm. In the Avrett table model (Avrett
\cite{avrett}), this size corresponds to the change in the plasma
frequency of $\sim$ 350 MHz. This very frequency separation
between ZS stripes was observed in Episode II (Fig.~\ref{lfig2}).

The large flare loop (green in Fig. \ref{lfig3}) extends for about
20$\arcsec$. Whistlers should go from the top to the base at a
group velocity of $\sim 10^9$ cm s$^{-1}$ for $\sim$ 0.7 s. It is
the longest duration of stripes in ZS fragments II and IV. The
shorter ZS episodes must have emanated from a smaller flare loop
(red in Fig. \ref{lfig3}). The Allen concentration model gives
much the same plasma frequency gradient (for example, see Fig,
IV.1 in Kruger \cite{kruger}). It corresponds to the density
height scale at these heights, about $10^7$ cm.

We can determine the magnetic field strength from the frequency
separation between stripes $\Delta f \sim$ 350 MHz. The whistler
group velocity peaks at the whistler frequency $f^{\omega} \sim
0.25 f_B$. With the value of separation between emission and
absorption taken as $\sim$ 0.5 $\Delta f$ and set equal to the
whistler frequency, it is easy to obtain the electron
gyrofrequency $f_B \sim$ 700 MHz (the magnetic field strength $B
\sim$ 250~G), which warrants our choice of the ratio
$\frac{{\omega}_B}{{\omega}_p} \sim$ 0.1.

We consider ZS pattern IV: two stripes with slightly wave-like
frequency drift (in zero value) and a large frequency separation,
$\sim$ 450 MHz. They are not identical, because the low-frequency
stripe started 0.4 second early, and their intensity and frequency
drift were not similar. Besides this the low-frequency stripe
bifurcated two times, exhibiting the frequency splitting at
02:13:03.4 and 02:13:03.9 UT. The sources must have been located
at the tops of different flare loops. Spike 6 (Fig. \ref{lfig2})
is in the spectrum at the beginning of the low-frequency stripe,
and according to the two-dimensional map, its source was at the
top of the large flare loop (green in Fig. \ref{lfig3}).

The absence of the frequency drift is likely to be tied to the
propagation of whistler packets almost at the level of equal
density at the top of the loop. In this case, whistlers may be
excited by the normal Doppler effect, or more precisely, the
scattering of fast particles by whistlers sets the distribution
function into the oscillation regime when excitation by normal or
anomalous Doppler effects is dominant. In a borderline case, both
the effects work simultaneously, and at these moments, we can see
the frequency splitting of ZS stripe because the excitation by
different effects occurs with a frequency shift. This excitation
may be referred to as the analog of the ``fan" instability (for
more detail, see Chernov \cite{chernov1996}). In the meter range,
this effect has been discussed twice to explain the frequency
separation of ZS stripes and the wave-like frequency drift
(Chernov et al. \cite{chernov1998}; Chernov \cite{chernov2005}).
The change in the frequency drift of stripes at the ``fan"
instability is associated with the smooth change in the whistler
group velocity direction. Episode III and the low-frequency ZS
stripe from IV might be tied to the small flare loop (red in Fig.
\ref{lfig3}).

The ZS episode V at higher frequencies (6.7--7.6 GHz) with a
frequency separation of $\sim$ 300 MHz and the explicit stripe
splitting occurred late in the phenomenon. It may be associated
with a lower loop arcade. The same is true of the short fragment
of ZS I, braided (in the terminology of Slottje \cite{Slottje81})
and of the very short episode of the small-scale ZS at 02:13:03.3
UT at $\sim$ 7.150 GHz: the frequency separation is $\sim$ 40 MHz;

The sporadic nature of the fine structure suggests multiple
pulsating acceleration of fast particles. However, the superfine
spike structure is not necessarily determined by a pulsating
acceleration as in Kuznetsov (\cite{kuznetsov2007}). It is much
more probable that the spike nature of the entire emission is
formed by the pulsating mechanism of emission. ZS is generated by
periodic whistler packets filling a magnetic trap. But whistlers
should occupy the entire radio source and propagate in different
directions. Spikes might therefore be related to the pulsating
interaction of whistlers with ion-sound waves and subsequent
coalescence with plasma waves. This mechanism is discussed in
Chernov et al. (\cite{chernov2001}). In the microwave range it may
be more effective because near the reconnection region the
presence of nonisothermic plasma ($T_e \gg T_i$) is much more
likely as a condition for ion sound excitation. ZS did not appear
at lower frequencies (2.6--3.8 GHz), and there was no spike
structure of the entire emission there. That is why we can say
that the spike structure appeared only in regions of the source
where whistlers existed.

However, at 5.2--7.6 GHz we could see many weak diffuse bursts
with different frequency drifts, which did not exhibit this spike
structure. Their polarization properties were similar to ZS ones.
It can even be said that they were the extensions of ZS. They may
therefore have been excited by the plasma mechanism at the upper
hybrid frequency, when there were no conditions for DPR.

Kuznetsov (\cite{kuznetsov2008}) proposes a new mechanism of
superfine structure in which the emission modulation is associated
with top-down propagation of MHD oscillations. But this mechanism
only gives the superfine structure for ZS stripes, not for the
whole emission, so the pulsating regime of whistlers remains
preferable in our case.

\subsection{New ZS models}

Searching for solutions to difficulties in different models,
recent papers have proposed several new mechanisms: the leakage of
Z-mode trapped into regular inhomogeneities of plasma density
(LaBelle et al. \cite{labelle}), and various versions of the
electromagnetic waves propagation through coronal inhomogeneities
(Laptukhov \& Chernov \cite{laptuhov2006}; Barta \& Karlicky
\cite{barta}; Ledenev et al. \cite{ledenev}; Laptukhov \& Chernov
\cite{laptuhov2009}). Two new exotic models have been proposed by
Fomichev et al. (\cite{fomichev}) (the explosive instability) and
by Kovalev (\cite{kovalev}) (periodic waves of electric charge).
All these new mechanisms must be realized only in specific
conditions. Their role in the ZS model hierarchy is still
ill-defined (Chernov \cite{chernov2010}).

\section{Conclusion}

We analyzed the unusual radio burst in the microwave range using
spectral data from Chinese spectrographs in the 5.2--7.6 GHz
frequency range and partially in the 2.6--3.8 GHz range and
spatial data from SSRT and Nobeyama radioheliographs. It proved to
be only the third case of a neat zebra structure appearing among
all observations at such high frequencies. Despite the short
duration of the burst (25 s), it provided a wealth of data for
studying the superfine structure with millisecond resolution. The
occurrence of the burst coincided with the formation of the
magnetic field ``delta"-configuration (for 3--5 hours) as a
precursor of the flare and with the appearance of local neutral
lines.

All the emission in the spectrum in the 5.2--7.6 GHz
       frequency range consisted of spikes lasting 5--10 ms
       in the instantaneous frequency band of 70 to 100 MHz.
These spikes make up the superfine structure of different
      drift bursts, fibers, and zebra-structure stripes.
      The zebra structure appeared about 3 s after the first
      spikes in the form of five patterns of stripes on different scales.
      The sources of some spikes were distributed along the flare
      loop.

The spikes were weakly polarized and had different signs.
      Their appearance concurred with the appearance of new strong
      R-polarized source at 17 GHz.
      The sizes of spike sources are estimated at $<$ 15$\arcsec$.
      Emission of spikes is generated during particle
      precipitation from the magnetic reconnection region down
      toward the footpoints of microloops. The spectral data point to an
      acceleration region height of $\sim$ 15000 km.

As the main mechanism for generating spikes
      (as initial emission) we suggest the coalescence of plasma
      waves with whistlers in a pulse regime of the interaction
      between whistlers and ion-sound waves. In this case, one
      can naturally explain the appearance of fibers and zebra
      structure with the same superfine structure in the form
      of millisecond pulses and the frequency splitting of
      zebra-structure stripes. Obviously some fast-drift
      bursts are excited by the beam instability in the pulse regime.

\begin{acknowledgements}
We are grateful to A.T. Altyntsev for his help in SSRT data
processing and for the fruitful discussion of all results in this
investigation. The authors thank B.L. Tan for valuable remarks on
the manuscript. We also appreciate the Nobeyama observatory team
for free access to the polarimeter and radioheliograph data.

This work was supported by RFBR (grants 11-02-00757, 08-02-92204-
GFNS, 09-02-92610-Ko, 09-02-00226, 10-02-00153-a) and by the
Program of RAS No. 16. Y.Y. is supported by NSFC (10921303) and
MOST (2011CB811401) grants. The  research  carried out by Dr.
Robert Sych at National Astronomical Observatories (NAOC) was
supported by the Chinese Academy of Sciences Visiting
Professorship for Senior International Scientists, grant No.
2010T2J24. This research was supported by a Marie Curie
International Research Staff Exchange Scheme Fellowship within the
7th European Community Framework Program.

\end{acknowledgements}

\end{document}